\documentclass[letterpaper, 10 pt, conference]{ieeeconf}  

\IEEEoverridecommandlockouts                              

\usepackage{amsmath}
\usepackage{subfigure}
\usepackage{algorithm}
\usepackage{multirow}
\usepackage{algpseudocode}
\usepackage{graphics} 
\usepackage{epsfig} 
\usepackage{mathptmx} 
\usepackage{times} 
\usepackage{amsmath} 
\usepackage{amssymb}  
\usepackage{xcolor}

\title{\LARGE \bf
Improving Low-Fidelity Models of Li-ion Batteries via Hybrid Sparse Identification of Nonlinear Dynamics
}

\author{Samuel Filgueira da Silva, Mehmet Fatih Ozkan, Faissal El Idrissi, Prashanth Ramesh
and Marcello Canova
\thanks{$^{1}$Department of Mechanical and Aerospace Engineering, Center for Automotive Research, The Ohio State University, Columbus, OH, USA. Emails: 
        {\tt\small filgueiradasilva.1@osu.edu, ozkan.25@osu.edu,  canova.1@osu.edu}}%
}

\begin{document}

\maketitle
\thispagestyle{empty}
\pagestyle{empty}

\begin{abstract}
Accurate modeling of lithium ion (li-ion) batteries is essential for enhancing the safety, and efficiency of electric vehicles and renewable energy systems. This paper presents a data-inspired approach for improving the fidelity of reduced-order li-ion battery models. The proposed method combines a Genetic Algorithm with Sequentially Thresholded Ridge Regression (GA-STRidge) to identify and compensate for discrepancies between a low-fidelity model (LFM) and data generated either from testing or a high-fidelity model (HFM). The hybrid model, combining physics-based and data-driven methods, is tested across different driving cycles to demonstrate the ability to significantly reduce the voltage prediction error compared to the baseline LFM, while preserving computational efficiency. The model robustness is also evaluated under various operating conditions, showing low prediction errors and high Pearson correlation coefficients for terminal voltage in unseen environments.
\end{abstract}

\section{INTRODUCTION}
Accurate models of li-ion battery (LiB) cells play an essential role in ensuring effective monitoring, diagnostics, and control of battery systems, which are vital to applications such as portable electronic devices, stationary energy storage, and electrified vehicle \cite{CHEN20194363,cano2018batteries}. Among various method, physics-based (electrochemical) models are extensively used for their ability to capture the complex electrochemical processes that govern the internal dynamics of LiBs \cite{ALI2024144360}. These models rely on complex nonlinear and partial differential equations (PDEs) to predict physical phenomena that lead to the cell terminal voltage, including ion and charge transport, equilibrium potentials and interfacial reaction kinetics. While highly accurate, the high computational cost associated with solving the constitutive equations makes real-time control and large-scale simulation challenging \cite{YU2023107661}.

This challenge motivates the development of reduced-order models, which offer faster computation time at the expense of accuracy, especially when capturing the intricate dynamics of battery physics at high C-rates \cite{WU2021137604}. To address this challenge, studies on data-driven modeling (DDM) techniques are being increasingly explored as a potential solution to complement traditional physics-based and reduced-order models \cite{9482997}. Moreover, methods based on machine learning have been extensively leveraged to predict different battery indicators such as state of charge (SOC) \cite{CHEMALI2018242,YANG2020117664,rodriguez2023discovering}, state of health (SOH) \cite{LI2020115340,LUCU2020101410}, and remaining useful life (RUL) \cite{9040661,ali2022hybrid}.

However, ensuring both robustness and interpretability in DDM for predicting battery dynamics remains a significant challenge, as these models often struggle to generalize across varying operating conditions and lack transparency on their mathematical structure, making it difficult to understand the rationale behind their decision-making process. In this context, regression techniques such as Sparse Identification of Nonlinear Dynamical Systems (SINDy) \cite{BRUNTON2016710} and Symbolic Regression \cite{udrescu2020ai,la2021contemporary} have emerged as promising approaches for generating interpretable models by identifying governing equations directly from data. For example, the authors in \cite{ahmadzadeh2023physics} employed SINDy with control (SINDYc) to predict the voltage and SOC dynamics of Lithium-ion batteries (LiBs). In their work, the training data set was generated using a Doyle-Fuller-Newman (DFN) electrochemical model. While the approach captures the highly nonlinear dynamics that govern the output voltage behavior, the main shortcoming lies in the reliance on manual processes for selecting the appropriate library terms and sparsification parameters. To address the limitations of SINDYc, recent research has introduced a sequentially thresholded ridge regression (STRidge) method that tunes the sparsification parameters \cite{AHMADZADEH2024110743}. However, this approach still relies on predefined library functions to construct the sparse data-driven model. If a large number of irrelevant or redundant functions are included in the library, this approach may fail to predict the output accurately. 

To overcome the aforementioned challenges, this study proposes a bi-level data-driven method integrating metaheuristic optimization with sparse regression. The proposed approach jointly optimizes the process of selecting both library terms and sparsification parameters to develop a more generalizable and parsimonious model. \par


 
The starting point of this study is a low-fidelity model (LFM) of a LiB cell that is obtained via model order reduction from an electrochemical model \cite{seals2022physics}. The prediction accuracy is improved by creating a data-driven model of the error dynamics via a Genetic Algorithm (GA)-guided STRidge (GA-STRidge) approach, which leverages data generated from a high-fidelity model (HFM) \cite{fan2016modeling}. By incorporating the proposed data-driven system identification, this study shows how the predictive accuracy of an LFM can be significantly improved while preserving computational efficiency and retaining knowledge of the effects that states and inputs of the LFM carry on the error dynamics.

The paper is structured as follows. Section II summarizes the equations of the LFM, while Section III describes the hybrid identification procedure. Finally, Section IV summarizes the training, validation and verification methodology.

\section{Overview of LiB Cell Model Equations}

\subsection{Overview of Li-ion Battery Models}
The starting point for this study is a low-fidelity model of a li-ion cell, developed originally in \cite{seals2022physics}. This model, known as the electrochemical equivalent circuit model (E-ECM), is obtained via model order reduction, linearization and approximation of the governing equations of an electrochemical model to achieve high computational efficiency while preserving the characteristic dynamics originating from the lithium transport in the positive and negative electrodes. 

This section summarizes the relevant equations of the LFM, including the approximations used. The terminal voltage $V_{LFM}$ is defined as follows:
\begin{equation} \label{eq:1}
\begin{split}
V_{LFM}(t) = U_p(c_{s,p},t) - U_n(c_{s,n},t) - (\eta_p(c_{s,p},t) + \\- \eta_n(c_{s,n},t) )+ \phi_{ohm}(t) - I(t)R_c
 \end{split}
\end{equation}

The kinetic overpotential $\eta_i$ and the exchange current density $i_{0i}$ are expressed as:
\textcolor{black}{\begin{equation} \label{eq:6}
\begin{split}
\eta_{i}(t) = \frac{\bar R T_0(-J_iI(t))}{Fi_{0,i}}
 \end{split}
\end{equation}}
\textcolor{black}{\begin{equation} \label{eq:5}
\begin{split}
i_{0i}(t) = exp\left( \left(  \frac{1}{T_{ref}} - \frac{1}{T(t)} \right) \frac{E_{ioi} }{\bar R}   \right) F k_i ~ \times \\  \sqrt{c_{s,i}(c_{s,max,i} - c_{s,i})c_{e,i}}
 \end{split}
\end{equation}}

\noindent where $i=p,n$ (positive or negative electrode), $E_{io,i}$ is the activation energy, $c_{max,i}$ refers to the maximum surface concentration, $c_{e,i}$ denotes the electrolyte concentration, and $T_{ref}$ is the reference temperature. A second-order Padé approximation is employed to compute the lithium surface concentration $c_{s,i}$ for the solid phase diffusion, resulting in a reduced order linear model:
\textcolor{black}{\begin{equation} \label{eq:7}
\begin{split}
c_{s,i}(s) = c_{s,i0} + (G_b(s) + G_d(s))\frac{-R_i}{3F\epsilon_{am,i}L_iA} I(s)
 \end{split}
\end{equation}}
where $G_b(s)$ represents the transfer function for bulk concentration and $G_d(s)$ refers to the transfer function for diffusion dynamics. These transfer functions are given by Eq. \eqref{eq:8}, containing physical parameters such as the diffusion coefficient $D_{i}$ and the particle radius $R_{i}$:
\begin{equation} \label{eq:8}
G_b(s) = \frac{\frac{2}{7}\frac{R_i}{D_i}s + \frac{3}{R_i}}{\frac{1}{35}\frac{R_i^2}{D_i}s^2 + s} \quad G_d(s) = \frac{\frac{R_i}{5D_i}}{\frac{1}{35}\frac{R_i^2}{D_i}s + 1}
\end{equation}
\par The potential loss in the electrolyte solution, $\phi_e$, is analogously obtained by a linear transfer function:

\textcolor{black}{\begin{equation} \label{eq:10}
\begin{split}
\phi_e(s) = (G_{pos}(s) + G_{neg}(s))\frac{C_1}{D_{e}}I(s)
 \end{split}
\end{equation}}

\noindent where $C_1$ is calculated as a function of the cell parameters (Eq. \ref{eq:13}), $G_{pos}$ and $G_{neg}$ (Eq. \ref{eq:11}) are the transfer functions describing the concentration dynamics of the electrolyte at the positive and negative current collectors:

\begin{figure*} [!h]
\par\noindent\rule{\dimexpr(0.5\textwidth-0.5\columnsep-0.4pt)}{0.4pt}%
\rule{0.4pt}{6pt}
\begin{equation} \label{eq:13}
\textcolor{black}{C_1 = 2RT\left(\frac{-L_{cell}}{A_s}\frac{1}{F^2c_{e0}}\right)(1-t_0^{+})(1+\beta)\left(0.601 -
0.24 \sqrt{\frac{c_{e0}}{1000}} +  0.982\left(1 - 0.0052(T_0-T_{ref}) \left(\frac{c_{e0}}{1000} \right)^{1.5}  \right)\right)}
\end{equation}
\vspace{\belowdisplayskip}\hfill\rule[-6pt]{0.4pt}{6.4pt}%
\rule{\dimexpr(0.5\textwidth-0.5\columnsep-1pt)}{0.4pt}
\end{figure*}
\begin{equation} \label{eq:11}
G_{\text{pos}}(s) = \frac{0.124\gamma_p}{\frac{0.1052L_{\text{cell}}^2}{D_e} s + 1}, \quad G_{\text{neg}}(s) = \frac{0.117\gamma_n}{\frac{0.0997L_{\text{cell}}^2}{D_e} s + 1}
\end{equation}

Finally, the potential drop caused by ohmic losses, $\phi_{ohm}$, is described by \eqref{eq:14}, where $\kappa$ depends on the initial electrolyte concentration $c_{e0}$ and cell temperature $T$.
\textcolor{black}{\begin{equation} \label{eq:14}
\begin{split}
\phi_{ohm}(t) = \frac{-I(t)L_{cell}}{\kappa A}
 \end{split}
\end{equation}}

Finally, the data used in this study for training, verification, and validation are generated using a high-fidelity model (HFM) as a virtual battery. Specifically, the extended single particle model (ESPM) is employed, as it accurately captures the key dynamics of Li-ion cells, including the solid-phase and electrolyte concentration behavior. For a detailed presentation of the equations governing the ESPM, as well as the underlying assumptions and processes, interested readers are referred to \cite{fan2016modeling}.

\section{Data-driven Identification Procedure}

\subsection{GA-STRidge Regression}
A data-driven identification procedure is employed to model the error dynamics between the LFM and HFM, $e_r$, calculated on the the terminal voltage:
\begin{equation} \label{eq:20}
e_r(t) = V_{HFM}(t) - V_{LFM}(t)
\end{equation}

The proposed method is a sparse learning via GA-guided STRidge regression, which aims to reconstruct a model that represents a dynamical functional relationship of the voltage error, current and LFM states, defined as follows: 
\begin{equation} \label{eq:21}
\mathbf{\hat{e}_r}[k+1] = \Theta(\mathbf{e_r}[k],\mathbf{I}[k],\mathbf{c_{s,p}}[k],\mathbf{c_{s,n}}[k])\xi
\end{equation}
\noindent where $\Theta \in \mathbb{R}^{n \times m}$ denotes the basis functions of the predicted model and $\xi \in \mathbb{R}^{m}$ is the sparse coefficient vector. This approach provides insights into how the error dynamics is influenced by the states of the LFM and the input (applied current $I(t)$), which is useful to investigate operating conditions under which the model discrepancies may escalate. This approach effectively augments the LFM to enhance its accuracy, while still significantly reducing the computational cost when compared to a HFM.\par
 The generation of the selected library function matrix $\Theta$ involves a systematic procedure where a metaheuristic algorithm based on GA is employed to select the most relevant basis functions from a broad set of potential candidate functions $\Theta_{lib}$ given by: 
\begin{equation} \label{eq:22}
\begin{split}
\Theta_{lib} = \{\Theta_{pol},\Theta_{cos},\Theta_{sin},\Theta_{tan},\Theta_{cosh},\Theta_{sinh},\Theta_{tanh}, \Theta_{ln}, \\ \Theta_{exp},\Theta_{sqrt}\}
\end{split}
\end{equation}
where the candidate set includes a diverse range of polynomial ($\Theta_{pol}$), trigonometric ($\Theta_{cos}$, $\Theta_{sin}$, $\Theta_{tan}$), hyperbolic ($\Theta_{cosh}$, $\Theta_{sinh}$, $\Theta_{tanh}$), logarithmic ($\Theta_{ln}$), exponential ($\Theta_{exp}$), and square root ($\Theta_{sqrt}$) functions. 



The GA-STRidge algorithm optimizes the selection process by iteratively evaluating the performance of different combinations of these basis functions, thereby identifying the subset $\Theta_S$ \eqref{eq:23} that best represents the underlying physics of the system:
\begin{equation} \label{eq:23}
\Theta_S = \left[ \begin{array}{c c c c c}
\vert & \vert &  & \vert & \vert \\
\Theta_{s,1} & \Theta_{s,2} & \cdots & \Theta_{s,m-1} & \Theta_{s,m} \\
\vert & \vert &  & \vert & \vert
\end{array} \right]
\end{equation}
This approach effectively mitigates the issue of poor performance that may be observed in sparse regression when a large number of irrelevant or redundant functions are included in the set of library functions \cite{AHMADZADEH2024110743}. By pre-selecting promising basis functions before applying the sparse regression, this procedure enhances higher interpretability and accuracy of the resulting model, ensuring that only the most promising terms are retained.

\par Once the library matrix $\Theta_S$ is defined, the STRidge regression technique is performed to calculate the coefficients $\hat{\xi}$, expressed as follows:
\begin{equation}  \label{eq:24}
\begin{split}
    \hat{\xi} = \arg \min_{\xi} \left\| \mathbf{\hat{e}_r}[k+1] - \Theta(\mathbf{U}[k],\mathbf{X}[k])\xi \right\|_{2} + \lambda_{1} \left\| \xi \right\|_{2} \\ \quad \text{s.t.:} \quad |\hat{\xi}_i| < \lambda_2
\end{split}
\end{equation}
\noindent where $\lambda_1$ represents the L2 regularization parameter that provides uniform coefficient shrinkage, distributing the effect of correlated variables across multiple features and avoiding model overfitting. The sparsification process is achieved by removing the unnecessary basis functions with absolute coefficient value below the specified threshold $\lambda_2$, as proposed by \cite{brunton2016discovering}.
In the evolutionary process, each candidate solution is assigned with a genome $P_i$ composed of a selected library matrix $\Theta_{S}$, and regularization $\lambda_1$ and sparsification $\lambda_2$ parameters. 
\begin{equation}
P_i = \{ \Theta_{S}, \lambda_1, \lambda_2 \}
\end{equation}
\par
\begin{figure}[h!] \label{fig:1}
    \begin{center}
       \includegraphics[angle=0,scale=0.38]{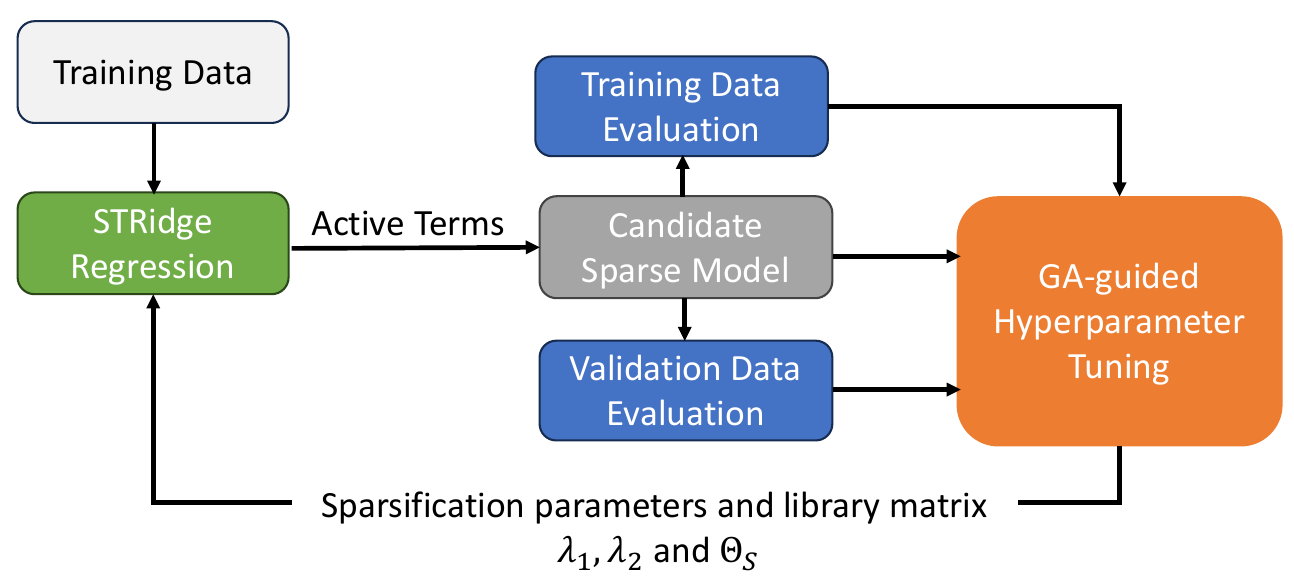}
    \caption{Schematic of the GA-STRidge algorithm.}
    \label{fig:schematic}
    \end{center}
\end{figure}

 The loss function in GA algorithm $\mathcal{L}$, expressed by:
\begin{equation} \label{eq:26}
\begin{split}
\mathcal{L}(\hat{\xi}, \Theta_S) = \alpha \left(
\frac{1}{n} \sum_{i=1}^{n} (\mathbf{\hat{e}_r} - \Theta_S\hat{\xi})^2 \right)_{train} + ~... \\ \beta \left(
\frac{1}{n} \sum_{i=1}^{n} (\mathbf{\hat{e}_r}- \Theta_S\hat{\xi})^2 \right)_{valid} + (1-\alpha-\beta) (N_{\xi}) 
\end{split}
\end{equation}
which is a combination of the weighted mean square error for the training set, validation set and number of active basis functions ($N_{\xi} = m$) to promote a parsimonious model that enhances interpretability and avoids overfitting, balancing accuracy and complexity. In this way, the validation error is incorporated into the loss function, enabling the model to be calibrated for a current profile not covered in the training process. The fitness function in the GA algorithm is then formulated as:
\begin{equation} \label{eq:27}
\begin{split}
   \{ \lambda_1, \lambda_2, \Theta_S \} = \arg \max_{\lambda_1, \lambda_2, \Theta} \left( 1 - \mathcal{L}(\hat{\xi}, \Theta_S) \right)  \\ \quad \text{s.t.:} \quad MSE(e_r, \hat{e_r})_{train} < \epsilon
\end{split}
\end{equation}
where $\epsilon$ is the minimum training error constraint in the STRidge regression to ensure a good fit to the training dataset.
The schematic of the GA-STRidge algorithm is presented in Fig. \ref{fig:schematic} and the overall procedure of the approach is described in Algorithm \ref{alg:bayesopt}.  
\begin{algorithm}
\caption{GA-guided Sparse Learning}\label{alg:bayesopt}
\begin{algorithmic}
\State $g \leftarrow 0$
\State Initialize population $(P(g=0))$
\State Evaluate population $(P(g=0))$
\While{(\textbf{not} stop-criterion)}
\State $P_{s}(g) \leftarrow selection(P(g))$
\State $P_{c}(g) \leftarrow crossover(P_s(g))$
\State $P_{m}(g) \leftarrow mutation(P_{c}(g))$
\State $P_{new}(g) \leftarrow [P_{c}(g),P_{m}(g)]$
\State \textit{STRidge Regression for each new candidate of $P_{new}(g)$}
\State \textit{Define $\lambda_1$, $\lambda_2$ and $\Theta$ based on $P_{new}(g)$}
    \State
$\hat{\xi} = \arg \min_{\xi} \left\| {e_r}[k] - \Theta(X[k], U[k])\xi \right\|_{2} + \lambda_{1} \left\| \xi \right\|_{2} $ 
    \For{$k$ from 1 to $iter$} 
    
        $sc = \{i: |\hat{\xi}_i| < \lambda_2\}$ (find small coefficients)
        
        $\hat{\xi}[sc] = 0$ (hard thresholding)

        $lc = \{j: |\hat{\xi}_j| \geq \lambda_2\}$ (find large coefficients)

        $\Theta_{lc} = \Theta[lc]$ (redefine library matrix)

        $\hat{\xi}[lc] = (\Theta^T_{lc}\Theta_{lc} + \lambda_1I)^{-1}\Theta^{-1}_{lc}e_r$ (update sparse vector)
        
    \EndFor
    \State Evaluate child solutions w.r.t training error
    \State Evaluate child solutions w.r.t validation error
    \State Evaluate the number of remaining functions
    \State Calculate loss function \eqref{eq:26}
    \State Evaluate child solutions w.r.t fitness value  \eqref{eq:27}
    \If{training criterion met} 
    
            \State $P(g) \leftarrow [P(g),  P_{new}(g)]$ (update population) 
            
    \EndIf
    \State $g \leftarrow g+1$
\EndWhile
\\
\Return{$\hat{\xi}_{opt}$ and $\Theta_{opt}$} that maximize fitness
\end{algorithmic}
\end{algorithm}



\section{Results and Discussion}
\subsection{Training and Validation Methodology}

The current profile in the training dataset is shown in Fig. \ref{fig:1}. The training current profile was obtained from simulating a battery electric vehicle (BEV) model across four distinct driving cycles: I) an ISO battery cycle with high C-rate; II) UDDS; III) HWFET and IV) US06. To prevent battery over-discharge, a constant current charge is applied between each profile. Additionally, the WLTC cycle is used to validate the model. The training and validation current profiles are applied in cascaded form to the LFM and HFM at 100\% initial SOC, corresponding to a voltage of 4.2V at equilibrium. For the testing, the Artemis Urban, JC08 and SC03 cycles are applied individually with 100\% starting SOC.

\begin{figure}[h!] \label{fig:1}
    \begin{center}
       \includegraphics[angle=0,scale=0.43]{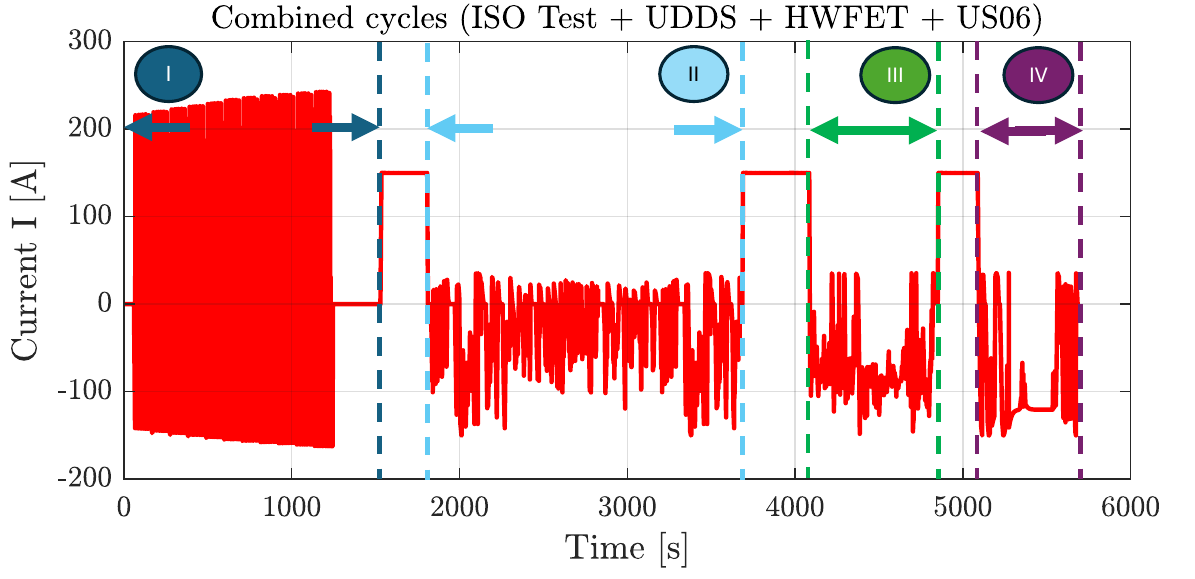}
    \caption{Current profile for training: combination of ISO battery cycle (I), scaled UDDS (II), HWFET (III) and US06 (IV)}
    \label{fig:1}
    \end{center}
\end{figure}



\begin{figure*}[!h] 
    \centering
    \subfigure[Voltage error dynamics in training and validation of GA-STRidge approach (zoom in error range of $\pm$ 0.1 V).]{{\includegraphics[angle=0,scale=0.5]{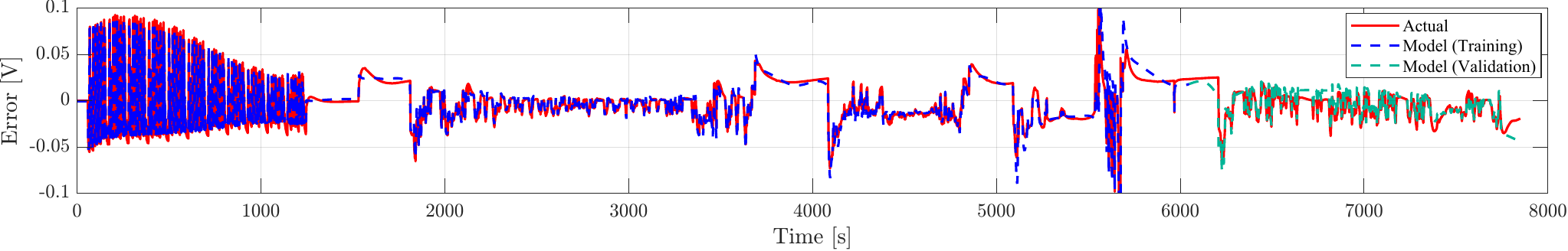}}} 
    \subfigure[Voltage prediction of HFM, LFM and hybrid model under training and validation tests.]{{\includegraphics[angle=0,scale=0.505]{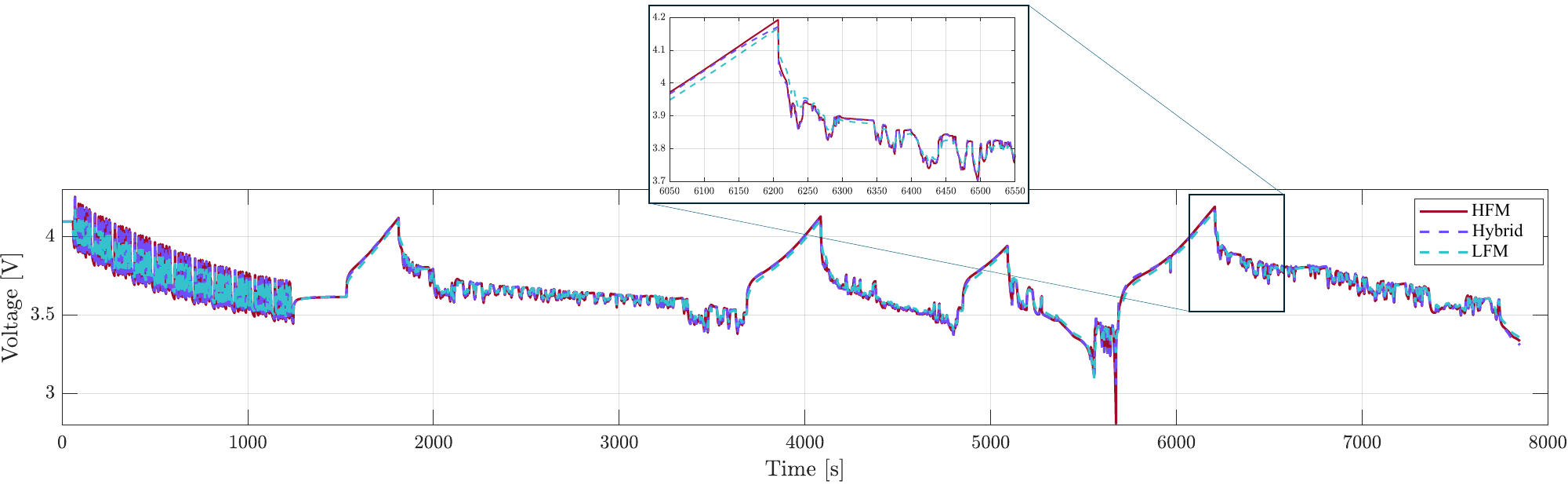}}}
    \caption{Predicted model performance under training and validation tests.}
    \label{fig:3}
\end{figure*} 


\subsection{Performance Evaluation}
Figure \ref{fig:3}a illustrates the performance of the hybrid model in predicting the error dynamics during training (blue dashed line) and validation (green dashed line). Although the model demonstrates a good fit to the training data, it effectively avoids overfitting by selecting optimal hyperparameters ($\lambda_1$ and $\lambda_2$) and a library matrix ($\Theta_S$) that promote parsimony. During training, the model achieves a mean squared error ($MSE$) of 6.99e-05 V$^2$ and a Pearson correlation coefficient ($\rho$) of 0.944. Additionally, the model effectively captures the error dynamics in the validation set and maintains robust performance when tested on unseen environments, as shown in Table \ref{table:1}. This is further supported by the voltage prediction accuracy of the augmented model, which outperforms the baseline LFM in both the training and validation set, as illustrated in Fig. \ref{fig:3}b.

\begin{table}[h!]
\centering
\caption{Model performance under validation and testing.}
\label{table:1}
\begin{tabular}{lcc}
\hline
\textbf{Cycle} & \textbf{MSE($e_r$, $\hat{e}_r$)} [V$^2$] & \textbf{$\rho$($e_r$, $\hat{e}_r$)}  \\ 
\hline
WLTC             & 4.886e-5 & \textcolor{black}{0.910}  \\
Artemis Urban    & 4.375e-5 & \textcolor{black}{0.951}  \\
JC08             & 1.097e-5 & \textcolor{black}{0.924}  \\
SC03             & 1.515e-5 & \textcolor{black}{0.921}  \\
\hline
\end{tabular}
\end{table}

\begin{table*}[h!]
\centering
\caption{Comparison of Voltage RMSE and Computational Time Reduction in Validation and Testing.}
\label{table:2}
\begin{tabular}{lccccc}
\hline
\multirow{2}{*}{\textbf{Cycle}} & \textbf{RMSE [V]} & \textbf{RMSE [V]} & \multicolumn{1}{c}{\multirow{2}{*}{\textbf{RRR [\%]}}} & \multicolumn{1}{c}{\textbf{Time Red. [\%]}} & \multicolumn{1}{c}{\textbf{Time Red. [\%] }} \\
               & \textbf{(HFM x LFM)} & \textbf{(HFM x Hybrid)} & & \textbf{(HFM x LFM)} & \textbf{(HFM x Hybrid)} \\
\hline
WLTC             & 0.0160 & \textcolor{black}{0.0070} & \textcolor{black}{56.28} & \textcolor{black}{94.70} & \textcolor{black}{55.97} \\
Artemis Urban    & 0.0141 & \textcolor{black}{0.0066} & \textcolor{black}{53.06} & \textcolor{black}{94.42} & \textcolor{black}{51.61} \\
JC08             & 0.0088 & \textcolor{black}{0.0033} & \textcolor{black}{62.51} & \textcolor{black}{93.22} & \textcolor{black}{59.03} \\
SC03             & 0.0117 & \textcolor{black}{0.0039} & \textcolor{black}{66.64} & \textcolor{black}{88.96} & \textcolor{black}{62.74} \\
\hline
\end{tabular}
\end{table*}


To evaluate the increase in model accuracy, the relative RMSE reduction (RRR) between the baseline LFM, the hybrid LFM and the HFM is calculated as follows:
\begin{equation}
\text{RRR} = \left(\frac{\text{RMSE}_{HFM,LFM} - \text{RMSE}_{HFM,Hybrid}}{\text{RMSE}_{HFM,LFM}}\right)100\%
\end{equation}

\par
Table \ref{table:2} presents a comparison of the voltage RMSE and computational time reduction between the HFM, baseline LFM, and hybrid models across different driving cycles. The results in the table demonstrate that the augmented LFM significantly outperforms the baseline LFM in terms of both accuracy and computational efficiency across all tested driving cycles. The RMSE values show that the hybrid model achieves a substantial reduction in voltage prediction errors, with improvements ranging from 53\% to 66\% depending on the cycle, indicating that the sparse learning process successfully captures the dynamics missing in the LFM. In terms of computational efficiency, the hybrid model offers significant time savings, reducing computational time by 51-62\% compared to the high-fidelity model, while presenting a slightly higher computation time than the low-fidelity model. These results highlight that the hybrid model strikes a balance between improving prediction accuracy and retaining high computational efficiency.

\subsection{Analysis of the Basis Functions in the Hybrid Model}
In this work, Singular Value Decomposition (SVD) is applied to analyze key patterns in the weighted basis functions $(\Theta\xi)$ derived by the GA-STRidge approach, allowing them to be ranked by importance. This ranking is essential for analyzing and identifying the most influential basis functions that capture the missing critical dynamics in the baseline LFM. SVD is particularly well-suited for this task, as it decomposes the functions into orthogonal components, with singular values used to quantify the significance of each function. By focusing on the basis functions with the highest singular values, the most important patterns in the system can be identified, leading to deeper insights into the underlying dynamics across various driving cycles and operating conditions. \par The SVD decomposes a rectangular matrix $\mathcal{S}$ into the product of three matrices, expressed as: 
\begin{equation} \label{eq:28}
\mathcal{S}_{(n \times m)} = U \Sigma V^{*}
\end{equation}
\noindent where $\mathcal{S}$ is matrix of the model features, with columns that represent each one of the model predictors computed at each instant of time. 
\begin{equation} \label{eq:29}
\mathcal{S}_{(n \times m)} = \begin{bmatrix}
    \Theta_{1,1}\xi_{1} & \cdots & \Theta_{1,m}\xi_{m} \\
    \vdots & \ddots & \vdots \\
     \Theta_{n,1}\xi_{1} & \cdots & \Theta_{n,m}\xi_{m}
\end{bmatrix}
\end{equation}

The columns of the $\mathcal{S}$ matrix can be expressed as a linear combination of the columns of $U$ (Eq. \eqref{eq:30}). By solving this linear square problem in Eq. \eqref{eq:32}, the coefficients $u_{i,k}$ are calculated, where $i$ and $k$ represent indexes for model predictors. 
\begin{equation} \label{eq:30}
\begin{aligned}
    \mathcal{S}_{1} &= u_{1,1} \overrightarrow{U_1} + u_{1,2} \overrightarrow{U_2} + \cdots + u_{1,m} \overrightarrow{U_{m}} \\
    & \vdots \\
    \mathcal{S}_{m} &= u_{m,1} \overrightarrow{U_1} + u_{m,2} \overrightarrow{U_2} + \cdots + u_{m,m} \overrightarrow{U_{m}}
\end{aligned}
\end{equation}
\begin{equation}  \label{eq:32}
\tilde{\mathbf{u}}_j = \arg\min_{\mathbf{u}_j \in \mathbb{R}^{m}} \| \mathcal{S}_j - \mathbf{u}_j \mathbf{U} \|_2^2, \quad \forall ~ j = 1, \ldots, m
\end{equation}
\par Hence, the features can be ranked based on a weighted average using singular values $\sigma_k$ as weights, as expressed by the equation below.
\begin{equation}
\bar{x}_{i} = \frac{\sum_{k=1}^{m} u_{i,k} \sigma_k}{\sum_{k=1}^{n} \sigma_k}, \quad \forall ~ i = 1, \ldots, m
\end{equation}
\par
\begin{figure}[h!] 
    \begin{center}
       \includegraphics[angle=0,scale=0.49]{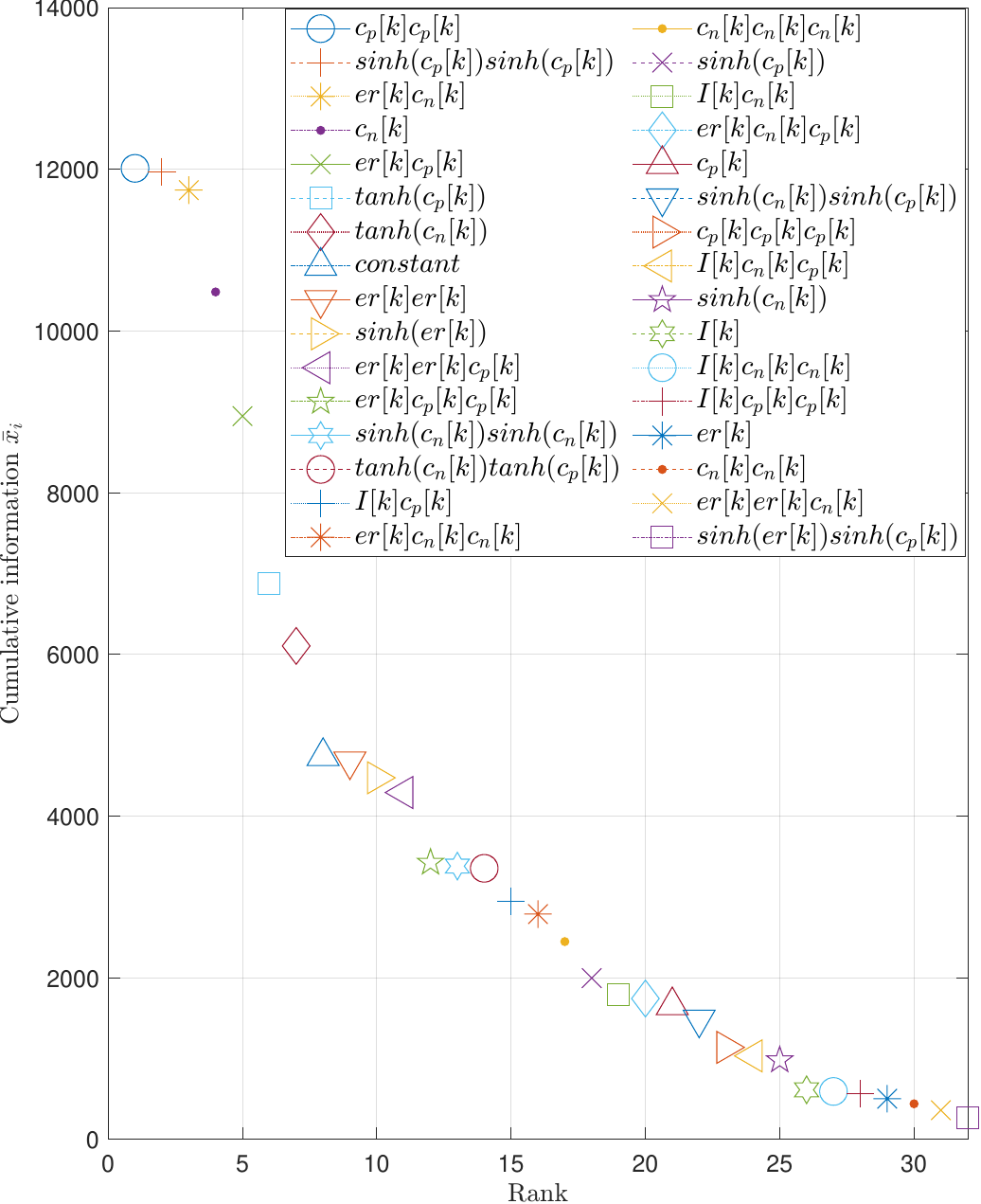}
    \caption{SVD ranking of the discovered basis function by GA-STRidge regression approach.}
    \label{fig:4}
    \end{center}
\end{figure}

 Figure \ref{fig:4} shows the ranking of the weighted basis functions. The vertical axis representing cumulative information $\bar{x}_{i}$ indicates how much each additional basis function contributes to the overall model's ability to predict the voltage error behavior. The model is composed of 81 features, of which 49 were eliminated via the sparse regression process, leaving 32 active basis functions.
It can be observed that solid phase concentrations appear in multiple high-ranking basis functions, suggesting these are critical in capturing the error dynamics. This can be explained by the different mathematical approximation methods that the LFM and HFM employ to model the solid phase diffusion, which significantly impacts the computation of the surface concentration and terminal voltage. In addition, the error state also plays an important role in reconstructing the model, especially when combined with other features like $c_p$ and $c_n$. This suggests that predictive accuracy is dependent on tracking the evolution of error over time. 

Another key observation is the prevalence of polynomial and hyperbolic functions in the analytical expressions of the identified model, despite the high variety of available functions the optimization algorithm can select, as shown in (\ref{eq:22}). Polynomial functions are crucial as they provide flexibility in approximating nonlinear relationships. The high importance of hyperbolic functions, on the other hand, can be attributed to their common appearance in PDE solutions. Such functions can represent exponential growth and decay behaviors (crucial for modeling processes like diffusion).

\section{Conclusions and Future Work}
This study demonstrates a data-driven approach for improving the accuracy of reduced-order electrochemical models for Li-ion batteries. By leveraging the GA-STRidge algorithm, the proposed hybrid model effectively compensates for the limitations of a low-fidelity model, significantly reducing prediction errors while maintaining computational efficiency. Tested across various driving cycles, the hybrid model shows robustness under different operating conditions, with reduced mean squared errors and improved correlation coefficients. 

Future work will focus on extending the method to identify governing equations from experimental data and comparing the performance with other model identification techniques.

\addtolength{\textheight}{-12cm}   




\section*{ACKNOWLEDGMENT}
The authors would like to acknowledge the Honda Research Institute for the inspiring discussions and helpful feedback that contributed to the research presented in this paper.
\section*{REFERENCES}

\begingroup
\renewcommand{\section}[2]{} 
\bibliographystyle{ieeetr}
\bibliography{mybibfile}

\begin{thebibliography}{10}

\bibitem{CHEN20194363}
W.~Chen, J.~Liang, Z.~Yang, and G.~Li, ``A review of lithium-ion battery for electric vehicle applications and beyond,'' {\em Energy Procedia}, vol.~158, pp.~4363--4368, 2019.
\newblock Innovative Solutions for Energy Transitions.

\bibitem{cano2018batteries}
Z.~P. Cano, D.~Banham, S.~Ye, A.~Hintennach, J.~Lu, M.~Fowler, and Z.~Chen, ``Batteries and fuel cells for emerging electric vehicle markets,'' {\em Nature Energy}, vol.~3, no.~4, pp.~279--289, 2018.

\bibitem{ALI2024144360}
H.~A.~A. Ali, L.~H. Raijmakers, K.~Chayambuka, D.~L. Danilov, P.~H. Notten, and R.-A. Eichel, ``A comparison between physics-based li-ion battery models,'' {\em Electrochimica Acta}, vol.~493, p.~144360, 2024.

\bibitem{YU2023107661}
H.~Yu, L.~Zhang, W.~Wang, K.~Yang, Z.~Zhang, X.~Liang, S.~Chen, S.~Yang, J.~Li, and X.~Liu, ``Lithium-ion battery multi-scale modeling coupled with simplified electrochemical model and kinetic monte carlo model,'' {\em iScience}, vol.~26, no.~9, p.~107661, 2023.

\bibitem{WU2021137604}
L.~Wu, K.~Liu, and H.~Pang, ``Evaluation and observability analysis of an improved reduced-order electrochemical model for lithium-ion battery,'' {\em Electrochimica Acta}, vol.~368, p.~137604, 2021.

\bibitem{9482997}
H.~Tu, S.~Moura, and H.~Fang, ``Integrating electrochemical modeling with machine learning for lithium-ion batteries,'' in {\em 2021 American Control Conference (ACC)}, pp.~4401--4407, 2021.

\bibitem{CHEMALI2018242}
E.~Chemali, P.~J. Kollmeyer, M.~Preindl, and A.~Emadi, ``State-of-charge estimation of li-ion batteries using deep neural networks: A machine learning approach,'' {\em Journal of Power Sources}, vol.~400, pp.~242--255, 2018.

\bibitem{YANG2020117664}
F.~Yang, S.~Zhang, W.~Li, and Q.~Miao, ``State-of-charge estimation of lithium-ion batteries using lstm and ukf,'' {\em Energy}, vol.~201, p.~117664, 2020.

\bibitem{rodriguez2023discovering}
R.~Rodriguez, O.~Ahmadzadeh, Y.~Wang, and D.~Soudbakhsh, ``Discovering governing equations of li-ion batteries pertaining state of charge using input-output data,'' in {\em 2023 American Control Conference (ACC)}, pp.~3081--3086, IEEE, 2023.

\bibitem{LI2020115340}
S.~Li, H.~He, C.~Su, and P.~Zhao, ``Data driven battery modeling and management method with aging phenomenon considered,'' {\em Applied Energy}, vol.~275, p.~115340, 2020.

\bibitem{LUCU2020101410}
M.~Lucu, E.~Martinez-Laserna, I.~Gandiaga, K.~Liu, H.~Camblong, W.~Widanage, and J.~Marco, ``Data-driven nonparametric li-ion battery ageing model aiming at learning from real operation data - part b: Cycling operation,'' {\em Journal of Energy Storage}, vol.~30, p.~101410, 2020.

\bibitem{9040661}
K.~Liu, Y.~Shang, Q.~Ouyang, and W.~D. Widanage, ``A data-driven approach with uncertainty quantification for predicting future capacities and remaining useful life of lithium-ion battery,'' {\em IEEE Transactions on Industrial Electronics}, vol.~68, no.~4, pp.~3170--3180, 2021.

\bibitem{ali2022hybrid}
M.~U. Ali, A.~Zafar, H.~Masood, K.~D. Kallu, M.~A. Khan, U.~Tariq, Y.~J. Kim, and B.~Chang, ``A hybrid data-driven approach for multistep ahead prediction of state of health and remaining useful life of lithium-ion batteries,'' {\em Computational Intelligence and Neuroscience}, vol.~2022, no.~1, p.~1575303, 2022.

\bibitem{BRUNTON2016710}
S.~L. Brunton, J.~L. Proctor, and J.~N. Kutz, ``Sparse identification of nonlinear dynamics with control ({SINDYc}),'' {\em IFAC-PapersOnLine}, vol.~49, no.~18, pp.~710--715, 2016.
\newblock 10th IFAC Symposium on Nonlinear Control Systems NOLCOS 2016.

\bibitem{udrescu2020ai}
S.-M. Udrescu and M.~Tegmark, ``Ai feynman: A physics-inspired method for symbolic regression,'' {\em Science Advances}, vol.~6, no.~16, p.~eaay2631, 2020.

\bibitem{la2021contemporary}
W.~La~Cava, B.~Burlacu, M.~Virgolin, M.~Kommenda, P.~Orzechowski, F.~O. de~Fran{\c{c}}a, Y.~Jin, and J.~H. Moore, ``Contemporary symbolic regression methods and their relative performance,'' {\em Advances in neural information processing systems}, vol.~2021, no.~DB1, p.~1, 2021.

\bibitem{ahmadzadeh2023physics}
O.~Ahmadzadeh, R.~Rodriguez, Y.~Wang, and D.~Soudbakhsh, ``A physics-inspired machine learning nonlinear model of li-ion batteries,'' in {\em 2023 American Control Conference (ACC)}, pp.~3087--3092, IEEE, 2023.

\bibitem{AHMADZADEH2024110743}
O.~Ahmadzadeh, Y.~Wang, and D.~Soudbakhsh, ``A data-driven framework for learning governing equations of li-ion batteries and co-estimating voltage and state-of-charge,'' {\em Journal of Energy Storage}, vol.~84, p.~110743, 2024.

\bibitem{seals2022physics}
D.~Seals, P.~Ramesh, M.~D'Arpino, and M.~Canova, ``Physics-based equivalent circuit model for lithium-ion cells via reduction and approximation of electrochemical model,'' {\em SAE International Journal of Advances and Current Practices in Mobility}, vol.~4, no.~2022-01-0701, pp.~1154--1165, 2022.

\bibitem{fan2016modeling}
G.~Fan, K.~Pan, M.~Canova, J.~Marcicki, and X.~G. Yang, ``Modeling of li-ion cells for fast simulation of high c-rate and low temperature operations,'' {\em Journal of The Electrochemical Society}, vol.~163, no.~5, p.~A666, 2016.

\bibitem{brunton2016discovering}
S.~L. Brunton, J.~L. Proctor, and J.~N. Kutz, ``Discovering governing equations from data by sparse identification of nonlinear dynamical systems,'' {\em Proceedings of the national academy of sciences}, vol.~113, no.~15, pp.~3932--3937, 2016.

\end{thebibliography}
\endgroup

\end{document}